\documentstyle[preprint,eqsecnum,aps]{revtex}
\begin{document}
\tightenlines
\begin{flushright} 

IFUP-TH 40/98
 \end{flushright}
\vskip 25pt plus 3pt minus 3pt

\begin{center}
  {\large \bf {Neutron Transfer to the Continuum Reactions}\footnote{Contribution to the
' Topical Conference on Giant Resonances ', Varenna, May 1998. To be published in
Nucl. Phys. {\bf A}}}

\end{center}
\vspace{.6em}
\begin{center}
{Angela Bonaccorso \\{\it Istituto Nazionale di Fisica Nucleare,
Sezione di Pisa, 56100 Pisa, Italy.\\ E-mail: BONACCORSO@PI.INFN.IT\\} }
\end{center}

\begin{abstract}

In this contribution we show that the theory of neutron transfer to the continuum reactions
is an useful tool to study different characteristics of the single particle structure of
nuclei.  In one example we discuss  properties of the single particle resonances in
$^{208}Pb$. Another interesting application deals with the neutron breakup from weakly bound
nuclei. Here one can use the theoretical calculations to help establishing the angular
momentum of the decaying state which is   experimentally not known.

\end{abstract}

%%%%%%%%%%%%%%%%%%%%%%%%%%%%%%%%%%%%%%%%%%%%%%% 

\section
{\bf Introduction}\label{int}

Transfer to the continuum reactions are 
quasi-elastic, peripheral heavy-ion reactions in which the   projectile of mass number  $A_P$,
transfers a nucleon to a target continuum final state of positive energy. The ejectile nucleus
of mass number $(A_P-1)$ is scattered at very
forward angles with almost the beam energy. This kind of reactions are relevant in the
medium-high beam energy domain where the incident energy per nucleon exceeds 
  the nucleon average binding
energy.  
 The mean energy loss is close to the incident energy per nucleon. Ejectile angular
distributions are  featureless while the inclusive ejectile energy loss spectra show
interesting features and by energy conservation are converted into residual nucleus
excitation energy spectra.  Projectile excitations are restricted to those leaving the
ejectile in an excited state below particle emission threshold since it is the $(A_P-1)$
nucleus which is detected.

The theory gives the following form of the neutron final energy $(\varepsilon_f)$ spectrum
\cite{bb} \begin{equation}{dP\over d\varepsilon_f}
\approx \Sigma_{l_f}(|1-\langle S_{l_f}\rangle |^2+1-|\langle
S_{l_f}\rangle |^2) B(l_f,l_i).  \label{dpde}\end{equation}

Eq.(1.1) can be easily transformed \cite{ab} into a neutron momentum
distribution  which is related to the measured ejectile momentum distribution. The physical
interpretation is that the projectile brings up the neutron which is scattered into a
continuum state by the target.  The  neutron-target interaction, in each partial wave
$l_f$, is represented by the optical model S-matrix for scattering of a free neutron by the
target nucleus.   The first term of
 Eq.(\ref{dpde}), proportional to
 $|1-\langle S_{l_f}\rangle |^2$, gives the neutron elastic breakup
spectrum in which the target is left into its ground state while the second term
proportional to the transmission coefficient $T=1-|\langle S_{l_f}\rangle |^2$ gives the
absorption spectrum \cite{bb}. 
 This term contains contributions from inelastic scattering of the
breakup neutron by the target nucleus and also from compound nucleus
formation.
The factor $B(l_f,l_i)$ is an elementary transfer
probability which depends on the details of the initial and final
states, on the energy of relative motion and on the distance of
closest approach $d$ between the two nuclei.  In
particular B contains as one of its factors the initial neutron parallel momentum
distribution in the projectile. The other factors depend  smoothly on the final neutron
energy in the case of a very weakly bound initial  state of low angular momentum (s or p).
In other cases the dependence can be quite important and it introduces strong distortion
effects on the initial momentum distribution.

 We obtain the final cross section, in the
strong absorption hypothesis for the relative motion  which is treated classically, by an
integration of Eq.(1.1) over all distances of closest approach $d$ larger than the strong
absorption radius \cite{bb,ab}.

 This paper is concerned with two applications of the theory of transfer
to the continuum reactions to the study of

- Damping of single particle resonance states in the continuum.

- Structure of weakly bound nuclei.

\section
{\bf Applications}\label{app}

The first example we wish to discuss is the reaction $^{208}Pb(^{40}Ar,^{39}Ar)^{209}Pb$ at
 $E_{inc}=41MeV/u$ whose inclusive energy spectrum
\cite{tiina} is given in Fig.(1). The full curve is the cross section calculated using
 Eq.(1.1). The close dotted curve is the elastic breakup obtained taking into account
only the first term of Eq.(1.1).
Absorption effects dominate in this reaction.

 We show also the energy distribution of the
 individual final angular momentum terms $l_f=8,9,10$ from the absorption part of
Eq.(1.1). Solid, dashed and dotdashed curves respectively. These terms give the cross
section for transfer to a particular final angular momentum state in the residual nucleus
and therefore represent the strength function for the corresponding single particle
state. The data shows a bump around 12 MeV  excitation  energy which is meanly due to the
$k_{17/2}$ state, but also another peak around 20 MeV which according to the calculation
originates from the $l_f=10$ state which is still not completely damped. The physical
reason for such a behavior is that these states have  high angular momentum, their high
centrifugal barriers keep them well inside the nuclear potential and there is little
mixing with low angular momenta states from which collective behavior and damping
usually originate. 
 A detailed discussion about the possibility of finding states with dominant single
particle characteristics in such a high excitation energy range, can be found in
\cite{an}.

This reaction has also another interesting characteristic, namely that there are several
possible  initial states contributing. Transfer from each of them would still leave the
ejectile in a state excited below particle threshold. Due to the finite experimental energy
resolution the data contains contributions from all of them. These initial states have different
angular momenta and therefore quite different initial momentum distributions. As a consequence
the shape of the spectra due to each of them are rather different, as shown in Fig.(2).
Solid line, dotted, dashed and close dotted refer respectively to $2s_{1/2}$,
$1d_{3/2}$, $2p_{3/2}$, $1f_{7/2}$ initial states. This is a very interesting characteristic of
transfer to the continuum reactions. Distortion effects caused by the neutron final state
interaction with the target are quite evident in this reaction \cite{tiina} because of the
dominant absorption effect of the neutron on the target.

In the case of weakly bound nuclei  breakup
dominates. Lately the interest on this class of processes has been renewed by the advent of
radioactive  beams\cite{ta} which are made up of unstable nuclei. These nuclei have one  or
two nucleons in a state of very weak binding energy. As a consequence the nucleon wave
function has an extended tail and one talks of 'halo  nuclei' \cite{hjj}.
In this case the reaction is a good probe of   the nucleon momentum
distribution in the projectile.  However spectra due to different initial angular momenta
have different widths and sometimes different shapes. 

For example the neutron breakup from weakly bound carbon isotopes has recently been
measured \cite{db}. These nuclei show interesting mixing of s-d orbits in their last bound
state wave functions but the exact amount of each component has not been established yet.
 Comparison of experimental data to theoretical calculation can help solving this problem.
We show  in Figs.(3) and (4) the neutron parallel momentum distribution from  the breakup of
$^{19}C$ and $^{17}C$ respectively on a $^{9}Be$ target. In the first case the calculation
from Eq.(1.1) reproduces the data if both s and d initial state contributions are included
with a dominant d (90\%) occupation probability. The s-contribution is responsible for
the narrow width at the top of the distribution but the tails of the distribution are due
to the d-component. In the $^{17}C$ case instead the breakup from the  d-state dominates
completely and the final momentum distribution is double peaked as the initial d
distribution shown in Fig.(5). In both Figs.(4) and (5)  the dotdashed line is the
d-distribution while the solid line is the s-distribution. However comparing the
distributions in Figs (4) and (5) we notice that the reaction mechanism has introduced
some distortion which can be seen in the data and it is reproduced by the calculation.
The origin of the distortion is in the mixing of the initial parallel and transverse
momentum distributions, induced by the reaction mechanism. We discuss this effect in a
forthcoming publication. It is more evident for   initial angular momenta higher than
$l=0,1$, partially because the experiment as well as the calculation make an
 average over the the initial angular momentum projections.

 Finally we mention that processes initiated by 
two-neutron halo nuclei are  still under debate as it can happen that one neutron breaks up
close to the interaction region with the target, while the other decays in flight \cite{bro}
due to a non negligible final state interaction with the projectile. This has been seen
clearly to  happen in the case of $^{6}He$ \cite{1}.

{\bf Acknowledgment}

This is the twentieth year since I started to collaborate with D. M. Brink. I wish to thank him
for his extraordinary example of clearness and rigor in Physics as well as in life.

{\bf Figure Captions}

Fig. (1). $^{208}Pb(^{40}Ar,^{39}Ar)^{209}Pb$ at 
$E_{inc}=41MeV/u$.
 
Fig. (2). Spectra for different initial states.

Fig. (3). Neutron parallel momentum distribution from $^{19}C$   breakup. 

Fig. (4). Neutron parallel momentum distribution from $^{17}C$   breakup. 

Fig. (5). Neutron momentum distribution for $l=0,2$.

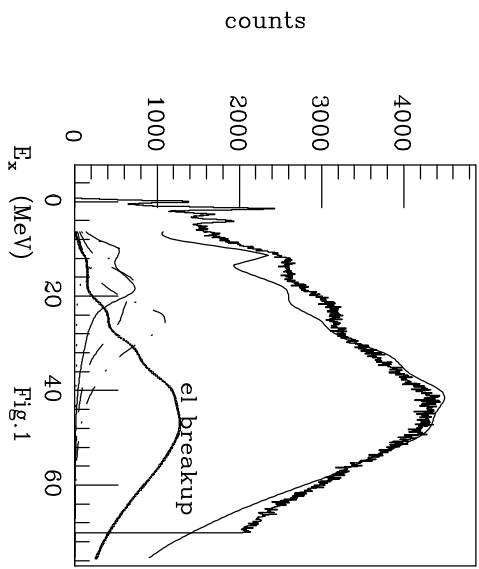
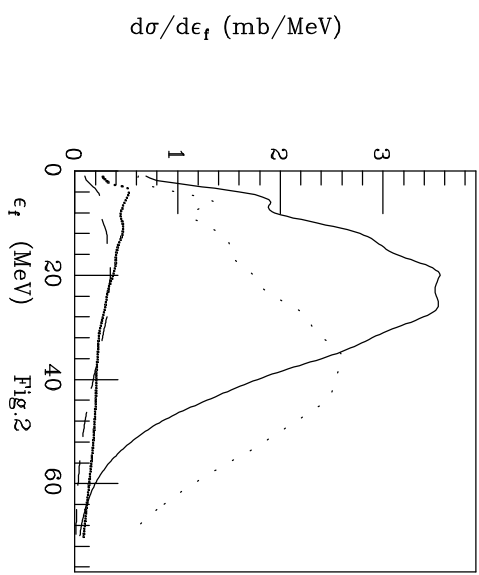
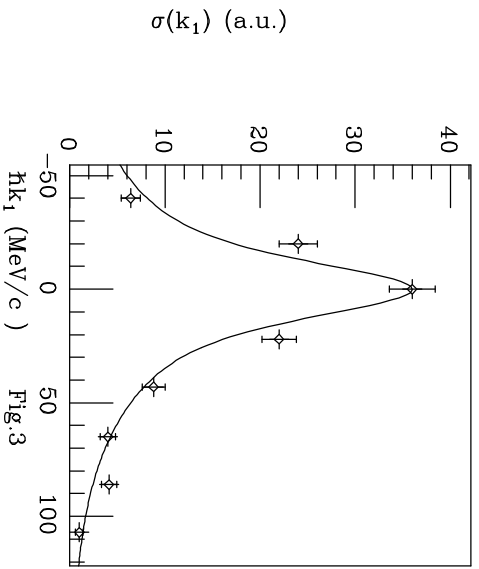
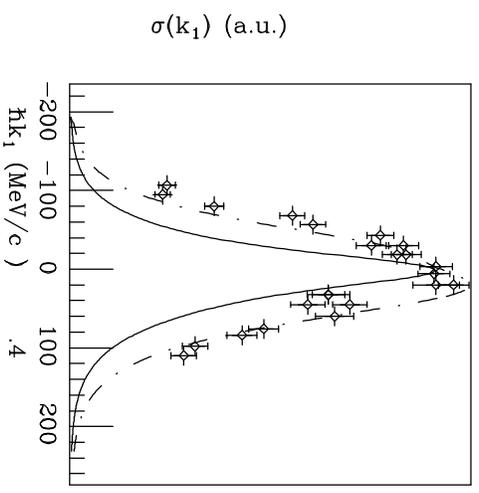
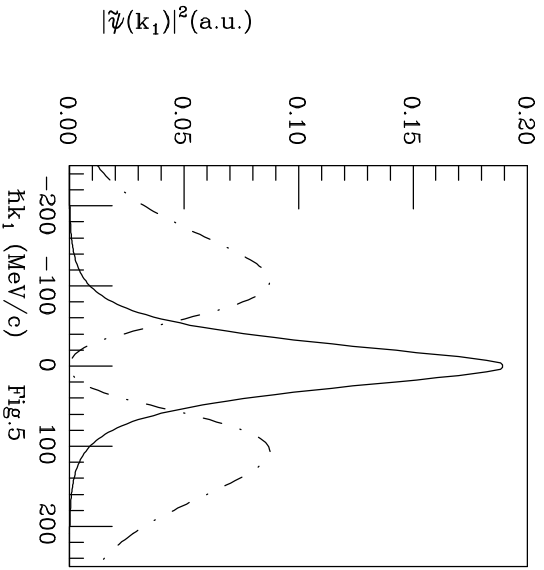
\end{document}